\documentclass[oldversion,letter]{aa}
\include{definitions}
\usepackage{graphicx}
\usepackage{txfonts}
\usepackage{ulem}
      \def\new#1 {{\bf #1 }}
      \def\cut#1 {\sout{#1} }

\begin{document}

\title{6.7\,GHz methanol absorption toward the Seyfert~2 galaxy NGC~3079\thanks{Based on observations with the 100\,m telescope of the MPIfR
 (Max-Planck-Institut f{\"u}r Radioastronomie) at Effelsberg.}}

\author{C. M. V. Impellizzeri \and C. Henkel \and A. L. Roy \and
   K. M. Menten }

\offprints{C. M. V. Impellizzeri,\\ \email violette@mpifr-bonn.mpg.de}
\institute{Max-Planck-Institut f\"ur Radioastronomie, Auf dem H\"ugel 69,
53121 Bonn, Germany }

\date{Received ; accepted  }

\abstract
{The detection of the 6.7\,GHz line of methanol (CH$_3$OH) is reported
for the first time toward an object beyond the Magellanic
Clouds. Using the Effelsberg 100\,m telescope, two absorption features
were identified toward the Seyfert~2 galaxy NGC\,3079. Both components
probably originated on lines-of-sight toward the central region,
presumably absorbing the radio continuum of the nuclear sources A, B,
and E of NGC~3079.  One absorption feature, at the systemic velocity,
is narrow and may arise from gas not related to the nuclear
environment of the galaxy. The weaker blue-shifted component is wider
and may trace outflowing gas. Total A-type CH$_3$OH column densities
are estimated to be between a few times 10$^{13}$ and a few times
10$^{15}$\,cm$^{-2}$. Because of a highly frequency-dependent
continuum background, the overall similarity of H{\sc i}, OH, and
CH$_3$OH absorption profiles hints at molecular clouds that cover the
entire area occupied by the nuclear radio continuum sources
($\sim$4\,pc).}

\keywords{galaxies: Seyfert -- galaxies: nucleus -- galaxies: starburst
-- galaxies: active -- galaxies: individual: NGC\,3079 -- radio lines:
galaxies}

\titlerunning{6.7\,GHz methanol beyond the Magellanic Clouds}
\authorrunning{ V. Impellizzeri et al.}

\maketitle

\section{Introduction}

The 5$_1$ $\rightarrow$ 6$_0$ A$^{+}$ transition of methanol
(CH$_3$OH) at 6.7\,GHz is one of the most prominent Galactic maser
lines (Menten 1991).  In our Galaxy, the line reaches flux densities
of up to several thousand Jy, not quite as much as the brightest
22\,GHz H$_2$O maser (e.g., Matveyenko et al. 2003) but exceeding the
flux densities of any known OH masers. The CH$_3$OH masers at 6.7\,GHz
are observed exclusively in star-forming regions, while OH masers near
1.7\,GHz are observed in the same regions and are often coincident on
subarcsecond scales (e.g., Menten et al. 1992). Hundreds of Galactic
6.7\,GHz methanol masers have been discovered since the early 1990s
(Pestalozzi et al. 2005). The 6.7\,GHz CH$_3$OH line is also found to
show absorption in certain regions; notably, deep absorption was found
toward our Galactic centre (Menten 1991).

{\it Thermal} emission from methanol at 96\,GHz was detected as early
as two decades ago toward two nearby external galaxies, NGC\,253 and
IC\,342 (Henkel et al. 1987). This detection, the large number of
luminous Galactic masers, and the existence of even more luminous
H$_2$O and OH ``megamasers'' (e.g., Lo 2005) have provided strong
motivation to search for 6.7\,GHz maser emission toward extragalactic
sources. Existing surveys targeted known OH megamaser galaxies and
objects with high infrared fluxes, but surprisingly, no detections
were obtained (Ellingsen et al. 1994a; Phillips et al. 1998; Darling
et al. 2003; Goldsmith et al. 2008). The only three detections
reported so far are from the Large Magellanic Cloud (Sinclair et al.
1992; Ellingsen et al. 1994b; Beasley et al.  1996). The intrinsic
brightness of these masers is similar to those of their stronger
Galactic counterparts.

Methanol masers form two distinct families. Class I masers are often
separated from the main source of excitation, whereas Class II masers
directly trace sites of high-mass star formation (Menten 1991). The
6.7\,GHz transition has become {\it the} Class~II maser line of choice
to study, but it requires intense emission from warm dust and
relatively cool gas to become inverted (Cragg et al. 2005). In regions
characterised by Class~I excitation, i.e., in the absence of a strong
far infrared (FIR) radiation field, the line is seen in absorption
(Menten 1991). Hence, if one intends to detect the 6.7\,GHz line,
instead of searching for maser emission, an alternative approach
involves absorption line studies against a strong background
continuum, tracing lines-of-sight toward active galactic nuclei (AGN).

Whilst all previous extragalactic methanol surveys at 6.7\,GHz have
been aiming at maser emission, the studies were also sensitive to
absorption. Nevertheless, none has been reported so far. We therefore
conducted a survey optimized for absorption and present the first
detection of 6.7\,GHz methanol absorption toward an extragalactic
source.

\section{Sample}

The sample of sources observed by us consists of eight Seyfert~2 or
LINER galaxies with a known high X-ray absorbing column ($N_{\rm H} >
10^{23}$\,cm$^{-2}$) and a radio continuum flux density $S_{\rm 6\,cm}
>$ 50\,mJy or a previous detection of a molecular absorption line
(NGC\,1052: Liszt \& Lucas 2004; NGC\,4261: Impellizzeri et al., in
preparation).

\section{Observations and data reduction}

We made spectroscopic observations, interspersed by continuum
measurements, with the 100\,m telescope of the MPIfR near Bonn,
Germany, during 2006 February to June and in 2007 March and
November. We used a dual-polarization HEMT receiver at the primary
focus.  System temperatures were 30--35\,K, corresponding to a system
equivalent flux density of about 22\,Jy. The full width at half power
(FWHP) beam size was 120\arcsec.

The observations were carried out in a position-switching mode,
integrating three minutes off- and three minutes on-source. We used
alternating $\pm$$\lambda$/8 focus shifts to eliminate standing waves
in the resulting spectra. The backend was the AK90 autocorrelator with
a total of eight bands, each consisting of 512 channels and covering
40\,MHz. The channel spacing was 78\,kHz, corresponding to
$\sim$3.5\,km\,s$^{-1}$. The pointing, obtained by cross scans toward
continuum sources, was accurate to about
10\arcsec--15\arcsec. Amplitude calibration was based on measurements
of the continuum emission of 3C\,48, assuming a flux density of
2.5\,Jy at 6.7\,GHz (Ott et al. 1994; see also Sect.\,4.2). The
absolute flux-density calibration is estimated to be accurate to
within $\pm$15\,\%. A third-order polynomial was fitted and subtracted
from the spectra to remove residual baseline ripples.

\section{Results}

\subsection{Extragalactic methanol}

Seven of the sources displayed in Table~\ref{table1} did not reveal
any spectral feature. In the case of absorption (see below), 5$\sigma$
upper limits on the line-to-continuum flux density ratio range from
$\sim$0.01 for the sources with strongest continuum (NGC\,4261,
NGC\,1052, NGC\,1068) to almost unity for the source with the weakest
continuum (NGC\,5135).

\begin{figure}
\centering
\includegraphics[angle=0,width=8.7cm]{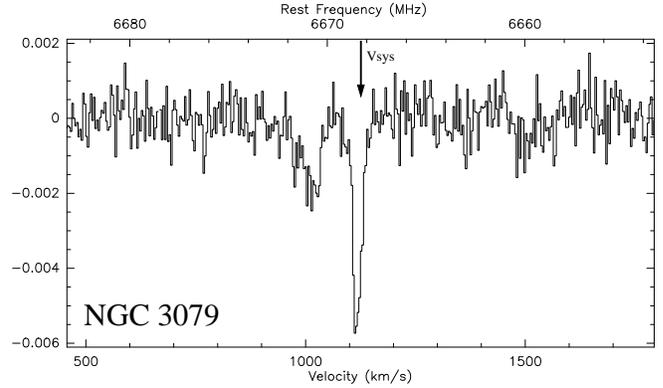}
\caption{5$_1$ $\rightarrow$ 6$_0$ A$^+$ methanol (CH$_3$OH) absorption
toward NGC\,3079 ($\alpha_{J2000}$ = 10$^{\rm h}$ 01$^{\rm m}$
57.8$^{\rm s}$, $\delta_{J2000}$ = +55$^{\circ}$ 40$^{\prime}$
47$^{\prime\prime}$) measured with Effelsberg. The ordinate gives flux
density in units of Jy. The channel width is 3.5\,km\,s$^{-1}$.}
\label{spectrum}
\end{figure}

We detected CH$_3$OH absorption toward NGC\,3079. The spectrum is
shown in Fig.\,\ref{spectrum}. Line parameters from Gaussian fits are
presented in Table~\ref{table2}. The profile shows a relatively strong
($\sim$6\,mJy) narrow component near the systemic velocity at $V_{\rm
systemic}$ = (1127$\pm$10)\,km\,s$^{-1}$ (Irwin \& Seaquist 1991), and
a weaker ($\sim$2\,mJy) broader blue-shifted component displaced by
$\sim$100\,km\,s$^{-1}$. While the FWHP linewidths of the two features
are quite different, the full width to zero power (FWZP) linewidths
{\it might} be similar ($\Delta V_0$ $\sim$
70\,km\,s$^{-1}$). Nevertheless, it is not possible to convincingly
disentangle a potentially broader systemic component from the dominant
narrow one.

\begin{table}
\caption{Observing parameters.}
\begin{tabular}{llcccc}
\hline\hline
Source      &  Date            & $t_{\rm int}$  &   $S_{\rm c}^{\rm a)}$ &  rms$^{\rm b)}$  \\
            &                  &  (min)         &     (mJy)        &  (mJy)  \\
\hline
NGC~3079 & combined spectrum         &  840   &   318            & 0.6  \\
NGC~4261 & 21 Feb, 2006              &  113   &  1007            & 2.2  \\
NGC~6240 & 25 May, 2006              &   75   &   117            & 2.0  \\
NGC~1052 & 16 Mar, 2006              &   62   &  1401            & 3.2  \\
NGC~1068 & 11 Jun, 2006              &   50   &  1385            & 5.4  \\
NGC~2110 & 16 Mar, 2006              &    5   &   130            & 8.0  \\
NGC~5506 & 26 May, 2006              &   26   &   155            & 3.8  \\
NGC~5135 & 17 Mar, 2006              &    3   &    48            & 8.0  \\
\hline
         &                           &        &                  &      \\
\multicolumn{5}{l}{a) 6.7\,GHz continuum flux density.}  \\
\multicolumn{5}{l}{b) Noise level for channel widths of 3.5\,km\,s$^{-1}$.}  \\
\end{tabular}
\label{table1}	
\end{table}

\subsection{The radio continuum of NGC\,3079}

For the first observations, taken in February 2006, our amplitude
calibration yielded a 6.7\,GHz continuum flux density of $S_{\rm c}$ =
(275$\pm$41)\,mJy. No suitable flux-density calibrator was observed in
March 2007, and $S_{\rm c}$ = (318$\pm$47)\,mJy was obtained in the
following November. For comparison, we obtain 260 and 315\,mJy by
extrapolating the 5\,GHz flux densities of Gregory et al. (1991) and
Mangum et al. (2008) with a spectral index of $\alpha$ = --0.7 ($S
\propto \nu^{\alpha}$). These agree within the uncertainities with the
measured fluxes and do not provide any evidence of continuum
variability.

\subsection{Apparent optical depths}

Observations of emission and absorption lines are
complementary. Emission commonly traces extended regions that show an
excitation that is significantly above the temperature of the cosmic
microwave background. To achieve this collisional excitation requires
that the density approximately matches or surpasses the critical
density of the line. Absorption lines have the advantage that
extremely tenuous gas can also be studied ($T_{\rm ex}$ $\sim$ 3\,K),
that the effective beam size is confined to the sometimes extremely
compact background continuum source(s), and that optical depths can be
obtained directly by a comparison of line and continuum flux
densities.

For the line profile shown in Fig.\,\ref{spectrum} and the line
parameters given in Table~\ref{table2} we calculate the optical depth,
$\tau$, using

$$
\tau = -{\rm ln}\,\left(\frac{-S_{\rm l}}{f_{\rm c} \times\ S_{\rm c}}\right).
$$

\noindent where S$_{\rm l}$ is the line flux density, S$_{\rm c}$  the
continuum flux density and $f_{\rm c}$ the source covering factor. For
the continuum background, we adopt $S_{\rm c}$ = (300$\pm$45)\,mJy
because most of the spectral data were obtained in November 2007 (see
Sect.\,4.2 for individual measurements). For the source covering
factor, we take $f_{\rm c}$ = 1, assuming that the absorber obscures
all of the detected radio continuum uniformly. While a uniform
coverage, i.e. the absence of small-scale clumping, is highly unlikely
(see the end of Sect. 5.2 for a rough estimate of $f_c$), this
nevertheless provides a firm lower limit to the true optical depth of
the bulk of the obscuring region.

\begin{table}
\caption{NGC\,3079 line parameters obtained from Gaussian fits}
\begin{centering}
\begin{tabular}{lccr}
\hline\hline
$\int{S {\rm d}V}^{\rm a)}$ & $V\,^{\rm b)}$ & \ \ $\Delta V_{1/2}\,^{\rm c)}$\ \
& \ \ \ \ \ \ $\tau(\rm CH_3OH)\,^{\rm d)}$     \\
(mJy\,km\,s$^{-1}$) & \multicolumn{2}{c}{(km\,s$^{-1}$)}            \\
\hline
--145$\pm$8 & 1117.9$\pm$0.6 & 24.2$\pm$1.6  &  0.0201$\pm$0.0046     \\
--115$\pm$12& 1009.9$\pm$3.0 & 57.4$\pm$7.4  &  0.0067$\pm$0.0018     \\
\hline
            &                &               &                        \\
\end{tabular}
\label{table2}
\end{centering}

a) The error does not include uncertainties due to calibration.  \\
b) Local standard of rest velocities following the optical definition. \\
c) Full width at half power linewidth.         \\
d) Uncertainties in the peak optical depths include calibration errors
in the line and continuum flux densities (Sects.\,3 and 4.2) and in
the Gaussian line fit. For the adopted continuum flux density, see
Sect.~4.3. Here, the source covering factor is assumed to be $f_{\rm
c}$ = 1. For an estimate of $f_c$ see the end of Sect.~5.2.
\end{table}

\section{Discussion}

\subsection{NGC\,3079: general properties}

NGC~3079 is an edge-on dusty spiral with a highly active
nucleus. Located at a distance of $\sim$15\,Mpc (de Vaucouleurs et
al. 1991), NGC~3079 hosts a nuclear starburst ($L_{\rm FIR}$ $\sim$
3$\times$10$^{10}$\,L$_{\rm \odot}$; Henkel et al. 1986) and a heavily
reddened active galactic nucleus (AGN) classified as Seyfert type~2
(Ford et al. 1986; Sosa-Brito et al. 2001). Observations with the
Hubble Space Telescope (HST) show narrow ionized filaments that arise
from the nuclear region above the plane of the galaxy to envelope a
1.3\,kpc sized superbubble (Cecil et al. 2001). Parts of this
superwind-blown structure are also traced by soft X-rays (Cecil et
al. 2002). A strong 6.4\,keV Fe K$\alpha$ line further supports the
presence of an AGN (Iyomoto et al. 2001; Cecil et al. 2002).

NGC~3079 also hosts a prominent H$_2$O megamaser (Henkel et al. 1984;
Haschick \& Baan 1985). The maser indicates the presence of a nuclear
disc of diameter $\sim$2\,pc, with the plane of the disc oriented
along the same north-south axis as the large-scale galactic and the
kpc-sized molecular disc seen in CO. From the rotation curve of the
masers, the enclosed mass within 0.4\,pc is $M_{\rm BH}$ $\sim$
2$\times$10$^6$\,$M_{\odot}$ (Trotter et al. 1998; Yamauchi et
al. 2004; Kondratko et al. 2005). NGC~3079 shows multiple radio
continuum components towards the nucleus (see below; Middelberg et
al. 2007) and also prominent jet-like protrusions originating in the
nuclear region (Duric \& Seaquist 1988; Baan \& Irwin 1995). These are
associated with the superwind and also extend toward the obscured back
side of the galaxy, which is not visible at optical wavelengths.

\subsection{Lines-of-sight toward NGC\,3079}

Which part of NGC\,3079 is absorbed by the methanol features shown in
Fig.\,\ref{spectrum}? It could be continuum emitted from the nuclear
region, from the jet-like protrusions, or from the large scale
disc. The line parameters of the main spectral feature, i.e. narrow
absorption near the systemic velocity, suggest that absorption occurs
in quiescent gas along the line-of-sight toward the nuclear region.
Here, velocity gradients of an edge-on galaxy should be minimal and
the gas should be found near $V_{\rm systemic}$. The small FWHP
linewidth ($\sim$24\,km\,s$^{-1}$; Table~\ref{table2}) also indicates
that the gas is probably not part of the nuclear region itself (cf.,
Hagiwara et al. 2004). To discuss the potentially much larger FWZP of
this feature (Sect.\,4.1), perhaps hinting at broad systemic
absorption, requires a spectrum with higher signal-to-noise ratio than
shown in Fig.\,\ref{spectrum}. The large FWHP linewidth of the
blue-shifted component is consistent with two interpretations. It may
either be caused by a high degree of turbulence near the nuclear
region, perhaps related to the expanding superbubble (Cecil et
al. 2001) or by differential rotation further out. Since the
approaching, blue-shifted sides of the nuclear disc, as well as the
kpc-sized molecular disc, are located north of the dynamical centre,
one might conclude that the gas responsible for this component may
also be found north of the nucleus.

To further constrain lines-of-sight, we should note that 6.7\,GHz
methanol is not the first radio line observed in absorption toward
NGC~3079. At lower frequencies, 1.4\,GHz H{\sc i} and 1.7\,GHz OH have
already been reported (Haschick \& Baan 1985; Baan \& Irwin 1995,
Sawada-Satoh et al. 2000; Hagiwara et al. 2004), while 4.8\,GHz
H$_2$CO absorption was not seen by Mangum et al. (2008; a weak signal
has been very recently detected, Mangum, priv.comm.). The H{\sc i} and
OH lines show the same two main components that are also observed in
the 6.7\,GHz CH$_3$OH line, again with the systemic feature being
stronger than the blue-shifted one. The H{\sc i} reveals a third
redshifted component at $V$$\sim$1250\,km\,s$^{-1}$. 22\,GHz H$_2$O
maser emission is quite different (see Hagiwara et al. 2002). While
extending from 890 to 1350\,km\,s$^{-1}$, strong emission is found
neither at the systemic nor at the blue-shifted velocity showing
CH$_3$OH and OH absorption. Instead, the main H$_2$O component is seen
at $\sim$ 960\,km\,s$^{-1}$. The H$_2$O 22\,GHz maser line requires
higher densities and kinetic temperatures than the other transitions
and apparently traces a different gas component.

Baan \& Irwin (1995) obtained interferometric H{\sc i} and OH maps
with 1\arcsec--2\arcsec\ resolution, and concluded that the lines
originate from the central $<$2\arcsec of the galaxy.  Hagiwara et al.
(2004) observed OH with subarcsecond resolution and suggest a nuclear
outflow as the origin for the blue-shifted component. In our Galaxy,
6.7\,GHz methanol is not as widespread as 1.4\,GHz H{\sc i} and
1.7\,GHz OH. Therefore, methanol absorption in NGC\,3079 is also most
likely restricted to the lines-of-sight toward the nuclear region.

Lineshapes of the lower frequency H{\sc i} and OH and the higher
frequency CH$_3$OH lines are similar but not identical. The H{\sc i}
and OH lines are broader ($\sim$100\,km\,s$^{-1}$). This may be caused
in part by the already mentioned, more widespread spatial
distributions of H{\sc i} and OH, but the radio continuum morphology
must also play a role. At 1.4\,GHz, 1.7\,GHz, and 6.7\,GHz, different
continuum components dominate. This is nicely illustrated by
Middelberg et al. (2007), who present a collection of radio continuum
maps between 1.7\,GHz and 22\,GHz. Sources E and F dominate at low
frequencies, while sources A and B, located about 20\,mas towards the
west, dominate at high frequencies. This implies that H{\sc i} and OH
trace lines-of-sight that are different from those traced by CH$_3$OH;
H{\sc i} and OH absorb against sources E and F and methanol absorption
should mainly arise toward sources A, B, and E. The similarity of the
spectral lineshapes requires the presence of extended molecular
complexes that are at least as extended as the distance between
sources B and F, $\sim$50\,mas or 4\,pc.

At 5\,GHz, continuum sources A, B, E, and F contribute only
75--80\,mJy or 25\% to 30\% to the total flux density (Middelberg et
al. 2007). If all of the methanol absorption is viewed against these
components, the {\it true} optical depths of the two absorption
features are about 3--4 times larger than the {\it apparent} optical
depths given in Table~\ref{table2}.

\subsection{Enhanced absorption in the 6.7\,GHz methanol line}

After having observed the $J_{\rm k}$ = 5$_1$ $\rightarrow$ 6$_0$
A$^+$ transition of A-type methanol and having estimated line
parameters and optical depths, physical and chemical implications
still have to be discussed. The lowest levels of the $k$=0 ladder of
A-type methanol have lower energies than those in the neighboring
k-ladders that are connected by allowed radiative transitions. As a
consequence, the energy levels of the $k$=0 ladder tend to be
overpopulated and, in the absence of a strong radiation field, one
expects enhanced absorption (also called ``over-cooling'' or
``anti-inversion'') in the $J_{\rm k}$ = 5$_1$ $\rightarrow$ 6$_0$
A$^+$ transition. This is verified by the statistical equilibrium
calculations of Walmsley et al. (1988) and Leurini et al. (2004) and
describes a Class I methanol maser environment.

The situation is dramatically different for sources in the vicinity of
an intense FIR field resulting from warm dust ($T_{\rm d}$ $>$ 100\,K)
heated by an embedded high-mass protostellar object surrounded by
cooler gas. Here, intense pumping, predominantly to the first
torsionally excited state via radiation around 30\,$\mu$m, and
subsequent decay determine the level populations and lead to strong
6.7\,GHz maser emission (e.g., Cragg et al. 2005).

The mere fact that we observed absorption and not emission already
tells us that pumping processes leading to widespread maser emission
are not the dominant excitation mechanism for methanol toward the
central region of NGC\,3079. While this does not exclude the presence
of 6.7\,GHz masers (a single 1000\,Jy maser at $D$=5\,kpc would show a
flux density of only 0.1\,mJy at the distance to NGC\,3079),
absorption dominates, presumably because it is more widespread. This
is remarkable because the nuclear region of NGC\,3079 appears to be
more active than that of our own Galactic centre region, where deep
absorption is also observed (Menten 1991).

With the measured peak apparent optical depths (Table~\ref{table2}),
with a source covering factor $f_{\rm c}$ $\sim$ 0.25--0.30 (end of
Sect.5.2), and in the absence of strong excitation by FIR photons, we
can estimate the methanol column density along the line-of-sight
toward NGC\,3079. There are, however, two important unknown
parameters, the kinetic temperature and the density of the gas. With
these values in the range of $T_{\rm kin}$ = 20--100\,K and $n$(H$_2$)
= 10$^3$--10$^6$\,cm$^{-3}$ and the Large Velocity Gradient code
kindly supplied by S. Leurini (see Leurini et al. 2004), we find that
A-type methanol column densities, $N$(A-CH$_3$OH), between
3$\times$10$^{13}$ and 6$\times$10$^{15}$\,cm$^{-2}$ produce
absorption at the observed levels.  The main uncertainty lies with the
density, with high densities yielding low column densities and vice
versa. Our calculations generally find excitation temperatures
$\ll$1\,K. Thus the 5$_1$ $\rightarrow$ 6$_0$ A$^+$ line indeed is
over-cooled.

\section{Summary and outlook}

We have detected the 6.7\,GHz transition of methanol, one of the most
prolific Galactic maser lines, for the first time in a source located
well beyond the Magellanic Clouds. The spectrum obtained toward the
starburst and H$_2$O megamaser galaxy NGC\,3079 shows two absorption
components with line-to-continuum ratios of about 1/50 and 1/150. This
is the signature of a Class I environment, where the absence of a
strong infrared radiation field inhibits the inversion of the level
populations. Instead, the line is characterised by ``anti-inversion''
or ``overcooling''.

Most of the absorption likely occurs toward the nuclear continuum
sources A, B, and E.  With no information on the density and kinetic
temperature of the gas, the A-methanol column density is poorly
defined and may range between a few times 10$^{13}$ and a few times
10$^{15}$\,cm$^{-2}$.

Interferometric measurements will be needed to reliably convert
``apparent'' into ``true'' optical depths. The presence of 6.7\,GHz
CH$_3$OH absorption not only toward the central region of the Galaxy
but also toward the more active spiral galaxy NGC\,3079 suggests that
such gas can be found in other extragalactic sources as well. To
reveal the physical parameters of the interstellar medium traced by the
6.7\,GHz line, the detection of additional methanol lines would be
desirable. Suitable candidates are the 2$_0-3_{-1}$ E (12.2\,GHz),
4$_{-1}-3_0$ E (36.2\,GHz), and 7$_0-6_1$ A$^+$ (44.1\,GHz)
transitions that should also be mapped, for comparison, toward the
Galactic centre region.

\begin{acknowledgements}

We wish to thank S. Leurini for the use of her CH$_3$OH Large Velocity
Gradient code, T. Krichbaum for useful discussions, and J. Mangum for
critically reading the manuscript.

\end{acknowledgements}

\end{document}